\documentclass[
reprint,
showpacs,
showkeys,
amsmath,amssymb,
aps,
]{revtex4-1}

\usepackage{graphicx}% Include figure files
\usepackage{dcolumn}% Align table columns on decimal point
\usepackage{bm}% bold math
\begin{document}

\preprint{APS/123-QED}

%\title{Detailed complex network community structure analysis \\ 
%using a selective eigenvector expansion of the adjacency matrix}

\title{Comprehensive spectral approach for \\ community structure analysis on complex networks}
%\title{New spectral approach for complex network structure analysis}

\author{Bogdan Danila}
\affiliation{BMCC, The City University of New York, 199 Chambers St, New York, New York 10007-1047}
\email{bdanila@bmcc.cuny.edu}

\date{\today}

\begin{abstract}
A simple but efficient spectral approach for analyzing the community structure of complex networks is introduced. It works the same way for all types of networks, by spectrally splitting the adjacency matrix into a ``unipartite" and a ``multipartite" component. These two matrices reveal the structure of the network from different perspectives and can be analyzed at different levels of detail. Their entries, or the entries of their lower-rank approximations, provide measures of the affinity or antagonism between the nodes that highlight the communities and the ``gateway" links that connect them together. An algorithm is then proposed to achieve the automatic assignment of the nodes to communities based on the information provided by either matrix. This algorithm naturally generates overlapping communities but can also be tuned to eliminate the overlaps.
\end{abstract}

 % Use showkeys class option to show keywords

\maketitle

%\tableofcontents

\section{Introduction}

Community structure detection has been one of the most important research topics in network science in recent years. Although no exact definition exists, a community is broadly understood as a set of nodes that ``work together to achieve a certain function of the network". It is usually assumed that there is a correlation between the density of connections and function, namely that subsets of the network whose nodes are more densely connected than in a random ``null model" are likely to perform some function together \cite{FortuRev, NewmanDivisive, NewmanPREspect, NewmanPNAS}. Alternatively, especially in the case of bipartite or directed networks, a frequently used assumption is that nodes that share many connections are likely to perform a common task \cite{FortuRev, GuimeraModBip}. The two assumptions have essentially the same meaning in the case of very densely connected communities, but are otherwise distinct. The method presented in this paper naturally identifies communities defined according to either assumption.

Various methods have been proposed so far to identify the community structure, most of them applying only to unipartite undirected networks \cite{FortuRev, NewmanBook, NewmanDivisive, NewmanGreedy, SimulAnn, ExtrOpt, NewmanPREspect, GraphPart1, GraphPart2, HierClust, PartClust, SpClShiMalik, SpClNg, SpectStoc1, SpectStoc2, SpectLap, WalkMod, LabelProp, Circuits, SpectralGong, SarkarSVD, NewmanEffPrin}. They include divisive algorithms \cite{NewmanDivisive}, graph partitioning \cite{GraphPart1, GraphPart1}, hierarchical clustering \cite{HierClust}, partitional clustering \cite{PartClust}, spectral clustering \cite{SpClShiMalik, SpClNg, SpectStoc1, SpectStoc2, SpectLap}, as well as more unusual methods \cite{WalkMod, LabelProp, Circuits}. However, the most commonly used methods are those based on the maximization of a goal function called modularity, introduced by Newman and Girvan \cite{NewmanGreedy, NewmanPREspect, NewmanPNAS}. The maximization is achieved using different heuristic approaches like greedy search \cite{NewmanGreedy}, extremal optimization \cite{ExtrOpt}, simulated annealing \cite{SimulAnn}, or spectral bisectioning \cite{NewmanPREspect, NewmanPNAS}. The latter has evolved into more sophisticated algorithms, which increase performance \cite{HoustonGang, SpectralGong, BasslerFastAccurate} or are specifically designed for bipartite networks \cite{GuimeraModBip, BarberModBip}, directed networks \cite{NewmanModDir}, or networks with overlapping communities \cite{GriechischOverl, LazarOverl, WangOverl, SzymanskiOverl}. Although community detection algorithms that use modularity as a goal function are known to suffer from a resolution problem which prevents them from detecting communities below a certain size \cite{LanciLim, ArenasLim, FortuLim, Decelle, Nadakuditi, RadicchiPRE, RadicchiEPL}, they are so far the most frequently used in the case of undirected networks with non-overlapping communities because modularity is based on a clear working definition of what it means for such a network to be modular \cite{FortuRev}. However, in the case of bipartite or directed networks and especially for networks with overlapping communities there is no universally accepted definition of modularity \cite{GuimeraModBip, BarberModBip, NewmanModDir, GriechischOverl, LazarOverl, WangOverl, SzymanskiOverl, FortuRev} and there is no way to directly compare the quality of partitions that have been obtained by maximizing different modularity functions. For this reason, it is important to have a community detection method that is independent of a definition of modularity, works the same way in all situations, and produces results compatible with modularity-based methods whenever comparison is meaningful.

The first steps in this direction were taken in Refs.~\cite{ArenasOptMap, SarkarSVD}. Although Ref.~\cite{ArenasOptMap} does not provide a method for identifying the community structure, it is notable for using a truncated singular value decomposition (SVD) of a ``contribution matrix" to analyze the structure of pre-determined communities and the relationship between them. The algorithm of Ref.~\cite{SarkarSVD} identifies the communities by using a singular value decomposition of the unsigned Laplacian matrix for unipartite networks, or of the rectangular adjacency sub-matrix for bipartite networks, followed by the application of a $k$-means clustering algorithm in the subspace spanned by the left and right singular vectors corresponding to the largest singular values. In this latter regard, they are still very close to the spectral clustering algorithms of Refs.~\cite{PartClust, SpClShiMalik, SpClNg, SpectStoc1}. Their algorithm has the drawback of using different matrices for uni- and bipartite networks and can only identify ``unipartite"-type communities (comprising nodes from both parties) on bipartite networks. In addition, Ref.~\cite{SarkarSVD} lacks a performance comparison with modularity-based methods in terms of ensemble averages. The community detection method introduced in this paper is simpler and works the same way for all types of networks. It starts by generating two matrices, in which ``unipartite" and respectively ``multipartite"-type communities (the latter consisting of nodes from a single party) are immediately visible. The entries of these matrices provide a measure of the affinity or antagonism between the different nodes which can be useful by itself (and likely sufficient for many purposes), but can also be used to generate either overlapping or non-overlapping community structures.

Finally, with the exception of \cite{SpectralGong}, all spectral algorithms proposed so far to maximize modularity perform recursive bisections of the network and its communities by using only the leading eigenvalue of the modularity matrix. The bisections must be combined with additional ``fine-tuning" \cite{NewmanPREspect, NewmanPNAS}, ``final tuning" \cite{HoustonGang} and possibly agglomeration \cite{BasslerFastAccurate} steps, without which the performance of these algorithms would be insufficient. These additional steps do not increase the complexity of the algorithms but require significant extra effort to program. A question of both theoretical and practical importance is whether a different type of spectral algorithm, that uses multiple eigenvectors of the adjacency matrix and is not specifically designed to maximize modularity, still needs such additional steps to achieve good performance. We present results showing that, except for extremely sparse or weakly modular networks, the algorithm proposed in this paper produces good to excellent community structures without additional steps.

\section{Method}

\subsection{Background}

Let $A$ be the adjacency matrix of a sparse network with $N$ nodes. There is no restriction on whether the network is uni- or bipartite, unweighted or weighted. In the weighted case, $A$ is understood to be the weights matrix. We will assume that the network is undirected, but directed networks can be represented as bipartite undirected ones for the purpose of community structure analysis \cite{GuimeraModBip}.

The goal is to partition the network into a set of communities $\{C_k\}$, with $k=\overline{1,K}$, that makes sense in light of the criteria mentioned in the first paragraph of the Introduction. Although the adjacency matrix is the most straightforward representation of a network, it has so far been considered unfit for the purpose of determining the community structure. The reason for this apparent inability and the way to deal with it are discussed in this section.

Community detection algorithms have been proposed that use either the stochastic matrix \cite{SpectStoc1, SpectStoc2} or different forms of the network Laplacian \cite{SpectLap, SarkarSVD}, but the most popular algorithms start with the definition of a modularity function. In the case of unipartite undirected networks, modularity is defined as
\begin{equation}
Q = \sum_{k=1}^K \sum_{i,j\in C_k} \left( A_{ij}-\frac{d_i d_j}{2m} \right) ,
\end{equation}

\noindent where $d_i$ is the degree of node $i$ and $2m=\sum_{i=1}^N d_i$. Modularity is then expressed as
\begin{equation}
Q = \frac{1}{2m}S^T M S ,
\end{equation}

\noindent where $M$ is the modularity matrix defined by
\begin{equation}
M_{ij} = A_{ij}-\frac{d_i d_j}{2m}
\end{equation}

\noindent and $S$ is a binary $N\times K$ matrix with $S_{ik}=1$ if node $i$ belongs to community $k$ and zero otherwise.

In the standard spectral bisectioning algorithm due to Newman \cite{NewmanPREspect, NewmanPNAS} as well as in its variants \cite{GuimeraModBip, BarberModBip, NewmanModDir, SzymanskiOverl}, $S$ is a column matrix and the network is recursively bisectioned according to the signs of the components of the eigenvector corresponding to the largest eigenvalue of the modularity matrix and then of its modified community-wide version until the modularity function can no longer be increased. There are also ``fine-tuning" \cite{NewmanPREspect, NewmanPNAS} and ``final tuning" \cite{HoustonGang} steps that can be added at the end of each bisection and at the end of the bisectioning process, respectively, to improve the performance of the algorithm.

Of particular interest are the variants introduced by Guimera \cite{GuimeraModBip} and Barber \cite{BarberModBip}, which are both specifically designed to deal with bipartite networks but detect different types of communities. The algorithm of Ref.~\cite{GuimeraModBip} finds communities that are subsets of only one party. Such communities will be called ``bipartite" or ``multipartite" in this paper. On the other hand, the algorithm described in Ref.~\cite{BarberModBip} finds cross-party communities, which will be called ``unipartite". As will be seen, the algorithm presented in this paper is capable of detecting both types of communities on bipartite and therefore also on directed networks.

In \cite{NewmanPREspect}, Newman points out the possibility of using more than one eigenvector of the modularity matrix but this idea has not been pursued until recently \cite{SpectralGong, SarkarSVD}. The algorithm proposed in Ref.~\cite{SpectralGong} uses orthonormal rotations in a space spanned by the eigenvectors corresponding to the $K$ largest eigenvalues of the modularity matrix while \cite{SarkarSVD} uses a singular value decomposition of the unsigned network Laplacian followed by $k$-means clustering in a similar space.

\subsection{General description}

On the other hand, it is obvious that the community structure can be regarded as a ``coarse-graining" of the network under analysis. The intuition behind the method proposed in this paper is to translate the coarse-graining algebraically into a representation of a community as a square sub-matrix whose entries are all positive or greater than a certain positive threshold, centered on the main diagonal of a simplified adjacency matrix. This makes sense if belonging to a community is viewed as being under the influence of a ``center of power", with all members interacting with each other through it. The problem of identifying the community structure (including the case of overlapping communities) then translates into finding all such sub-matrices that are maximal (not contained within larger ones).

\begin{figure}
\scalebox{0.35}[0.35]{\includegraphics{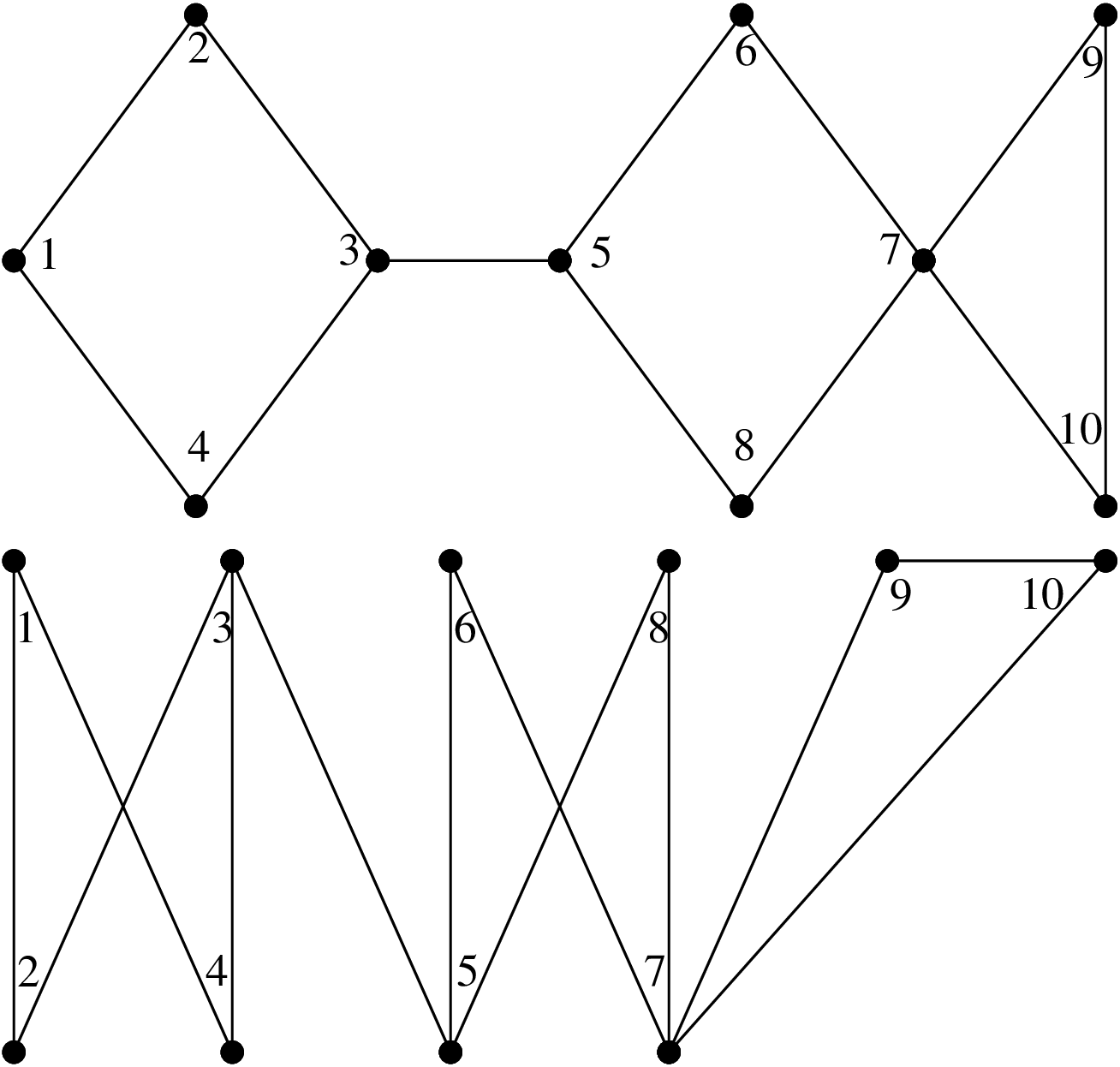}}
\caption{\label{fig:net10pics} A simple nearly-bipartite network.}
\end{figure}

Sub-matrices of the kind described above are nowhere to be found in the adjacency matrices of typical real-world or model networks. Networks composed of sparsely interconnected cliques come closest to this picture but even they have all diagonal elements equal to zero unless self-loops are allowed. In order to obtain a coarse-grained version of the adjacency matrix it seems natural to perform a singular value decomposition $A=U\Sigma V^T$ \cite{GolubBook} and then retain only the terms corresponding to the largest $K<N$ singular values,
\begin{equation}
A_{\{1-K\}} = \sum_{k=1}^K \sigma_k U_{:k} V_{:k}^T .
\end{equation}
\noindent Here $U$ and $V$ are orthogonal matrices whose columns are the left and right singular vectors of matrix $A$ while $\Sigma$ is diagonal with non-negative entries $\sigma_k$. This is reminiscent of approaches used in some lossy image compression and face recognition algorithms as well as of the principal component analysis method used in statistics \cite{ArenasOptMap, SarkarSVD}. A low-rank approximation of the adjacency matrix is expected to retain only its most important features, enhancing sets of similar rows or columns, introducing additional links within the densely connected subsets, and weakening the links between them \cite{GolubBook}. This is exactly what is needed in order to reveal communities defined either by high density of links or by similarity of connection, as discussed in the first paragraph of the Introduction. Moreover, it is known that retaining the first $K$ singular values from an SVD leads to the best rank-$K$ approximation of the original matrix in terms of Frobenius norm \cite{GolubBook}. Everything seems right, and yet, if the method is applied as described above, it gives fair results on some networks but completely fails to identify a meaningful community structure on many.

A simple example is the network shown in Fig.~\ref{fig:net10pics}, which is nearly bipartite except for the link between nodes 9 and 10. The network is shown in two different layouts, which emphasize the unipartite and bipartite communities respectively. The first term of the expansion in Eq.~(4) does contain information about the relative importance of the nodes within the network, which is not surprising, since $U_{:1}=V_{:1}$ defines the eigenvector centrality measure. As more terms are added, though, the singular value expansion simply converges towards the adjacency matrix without ever revealing a community structure.

% The fact that the adjacency matrix seems unfit for community detection has been noted before and  all methods introduced so far avoid it. In the following we show that it is possible to obtain a meaningful community structure directly from the adjacency matrix and that, in addition, two types of community structure, similar to those described in Refs.~\cite{GuimeraModBip} and \cite{BarberModBip} can be obtained in the case of bipartite or directed networks.

\begin{figure}
\scalebox{0.35}[0.35]{\includegraphics{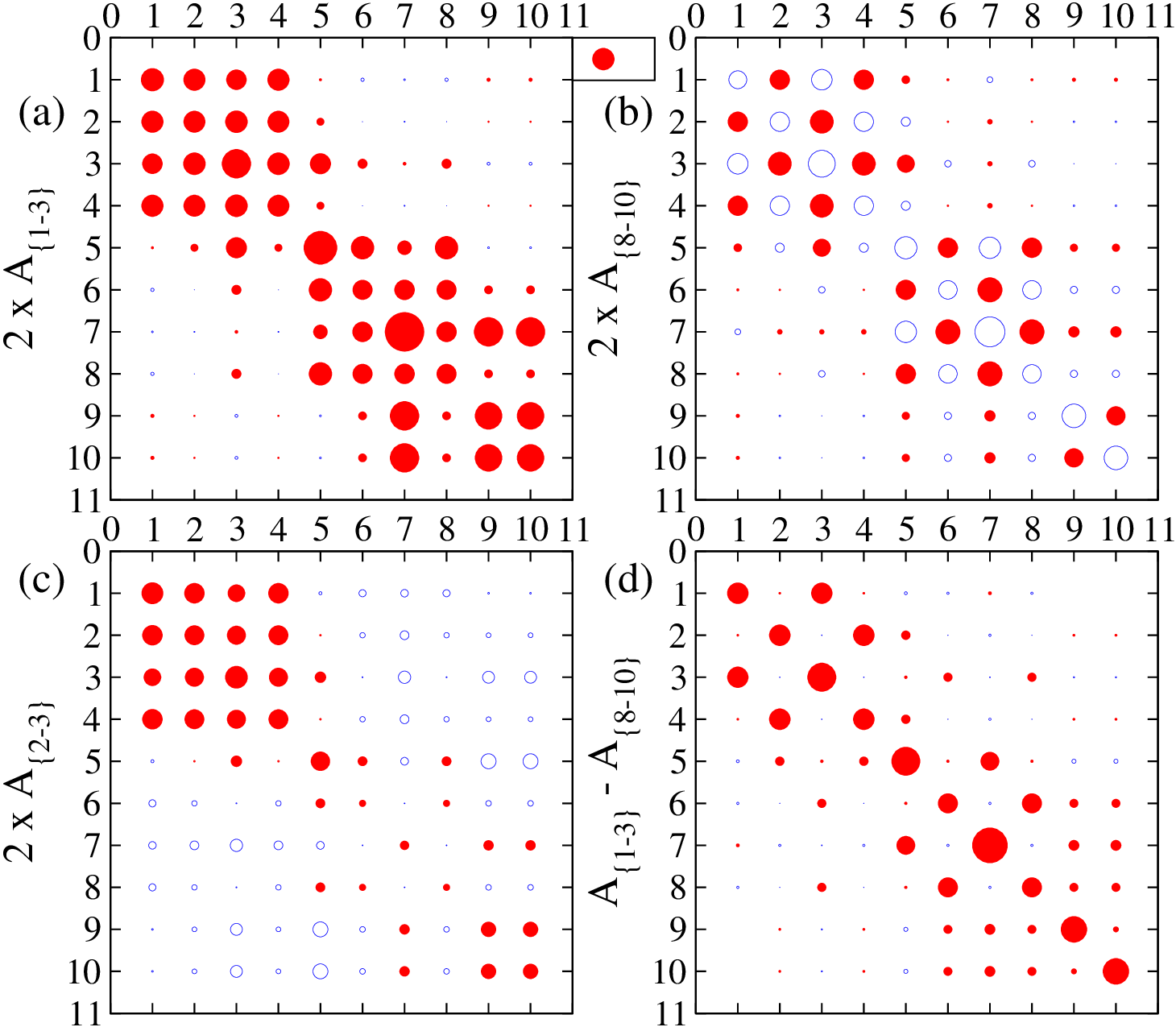}}
\caption{\label{fig:net10res} (Color online) Split eigenvalue expansions of the adjacency matrix for the network in Fig.~\ref{fig:net10pics}. Red (solid) and blue (hollow) dots represent positive and negative matrix entries, respectively. The dot in the legend box has unity diameter.}
\end{figure}

To understand the root of the problem, note first that for real symmetric matrices the singular value decomposition is closely related to the eigenvalue decomposition $A=U\Lambda U^T$: the singular values are the absolute values of the eigenvalues, $\sigma_i=|\lambda_i|$, and any negative eigenvalue signs are transferred to the columns of $U$ on the right to form $V$. Retaining the largest $K$ singular values in an SVD is the same as retaining the largest $K$ eigenvalues {\it in absolute value}. However, individual rank-1 terms of the form $\lambda_i U_{:i} U_{:i}^T$ in the eigenvalue expansion of $A$ tell different stories when interpreted in terms of community structure depending on the sign of $\lambda_i$.

If $\lambda_i>0$, the matrix has two blocks with positive entries on the main diagonal and two off-diagonal blocks with negative entries. This corresponds to a partition of the network into two unipartite-style communities, with the positive matrix elements quantifying affinity and the negative ones quantifying antagonism between the nodes.

If $\lambda_i<0$, the blocks with positive entries are off-diagonal, which corresponds to a bipartite approximation of the network, with two same-party communities appearing in the negative blocks and the connections between the nodes in the positive ones. This is reminiscent of Newman's observation \cite{NewmanPREspect} that the eigenvector corresponding to the largest negative eigenvalue of the modularity matrix $M$ can be used discern a (nearly-)bipartite structure.

\begin{figure}
\scalebox{0.35}[0.35]{\includegraphics{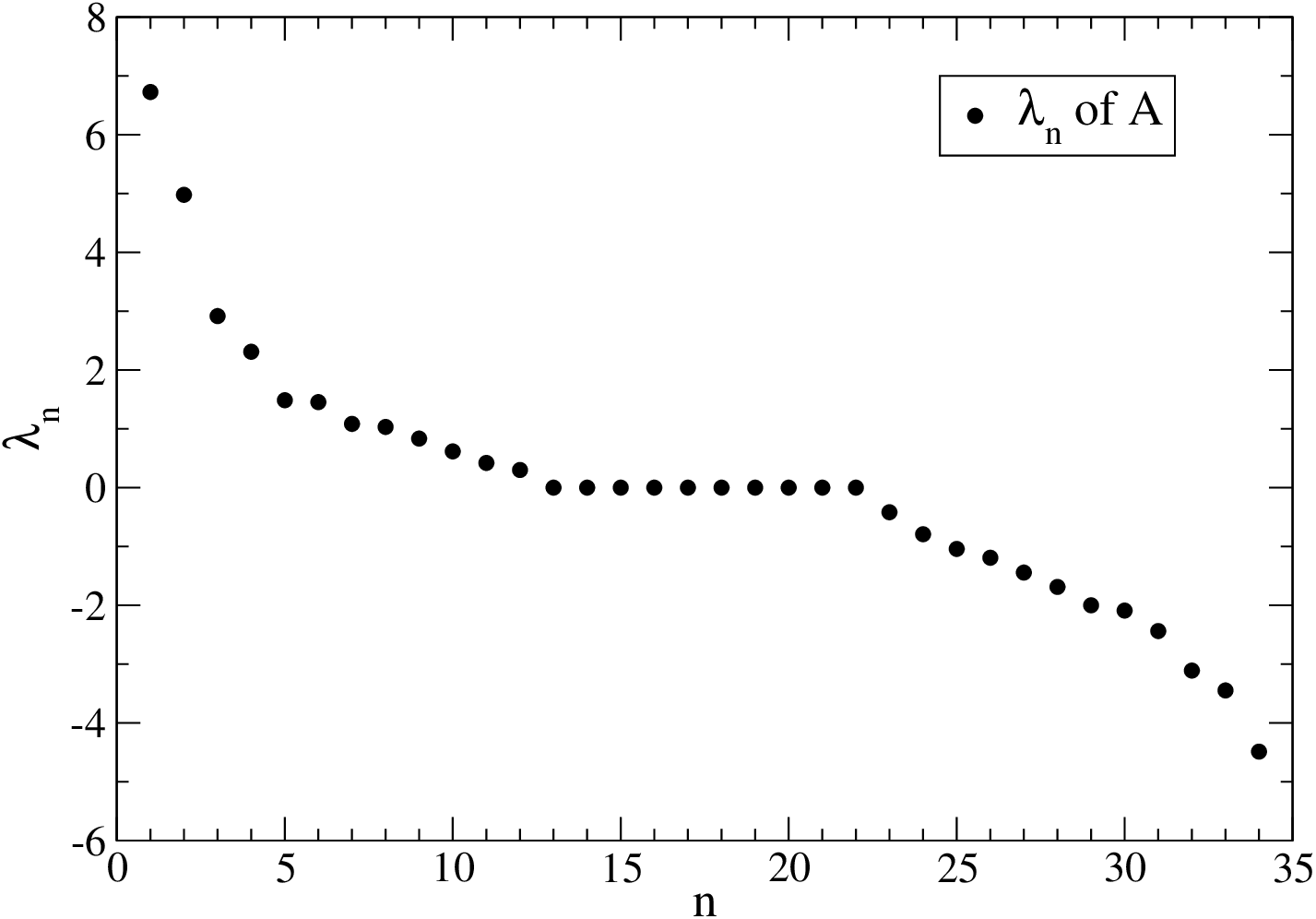}}
\caption{\label{fig:EWkarate} The eigenvalues for Zachary's karate network. Prominent positive eigenvalues 1 through 4 define the unipartite community structure. Prominent negative eigenvalue 34 defines a bipartite approximation of the network.}
\end{figure}

It is known \cite{SpectGraph} that bipartite networks have symmetric positive and negative eigenvalues of the adjacency matrix. In addition, many unipartite networks have large negative eigenvalues, of magnitude comparable to the largest positive ones. This means that two mutually exclusive types of community description interfere if one simply performs a singular value decomposition of the adjacency matrix. The key to correctly revealing the community structure of a network based on the adjacency matrix is to spectrally split it into an ``unipartite" and a ``multipartite" component, the former constructed using exclusively the eigenvectors with positive eigenvalues and the latter the eigenvectors with negative eigenvalues,
\begin{eqnarray}
A_U &=& \sum_{\lambda_k>0} \lambda_k U_{:k} U_{:k}^T \\
A_M &=& \sum_{\lambda_k<0} \lambda_k U_{:k} U_{:k}^T .
\end{eqnarray}

For the purpose of revealing the community structure, we can retain the largest $K_p$ positive eigenvalues and the largest $N-K_n+1$ negative eigenvalues. Assuming the eigenvalues are listed in decreasing order, the ``coarse-grained" versions of these matrices are
\begin{eqnarray}
A_{\{1-K_p\}} &=& \sum_{k=1}^{K_p} \lambda_k U_{:k} U_{:k}^T \\
A_{\{K_n-N\}} &=& \sum_{k=K_n}^{N} \lambda_k U_{:k} U_{:k}^T .
\end{eqnarray}

The results of such a spectral split for the network in Fig.~\ref{fig:net10pics} are shown in Figs.~\ref{fig:net10res} (a) and (b). The first matrix reveals communities in ``unipartite" mode: nodes from one party that are densely connected as second-order neighbors are lumped together with the first-order neighbors through which they are connected into cross-party communities. The \emph{negative} entries of the second matrix reveal communities in ``bipartite" mode, with nodes from only one party that share neighbors in the other lumped by themselves. The results for this network are discussed in more detail in subsection E.

\begin{figure}
\scalebox{0.35}[0.35]{\includegraphics{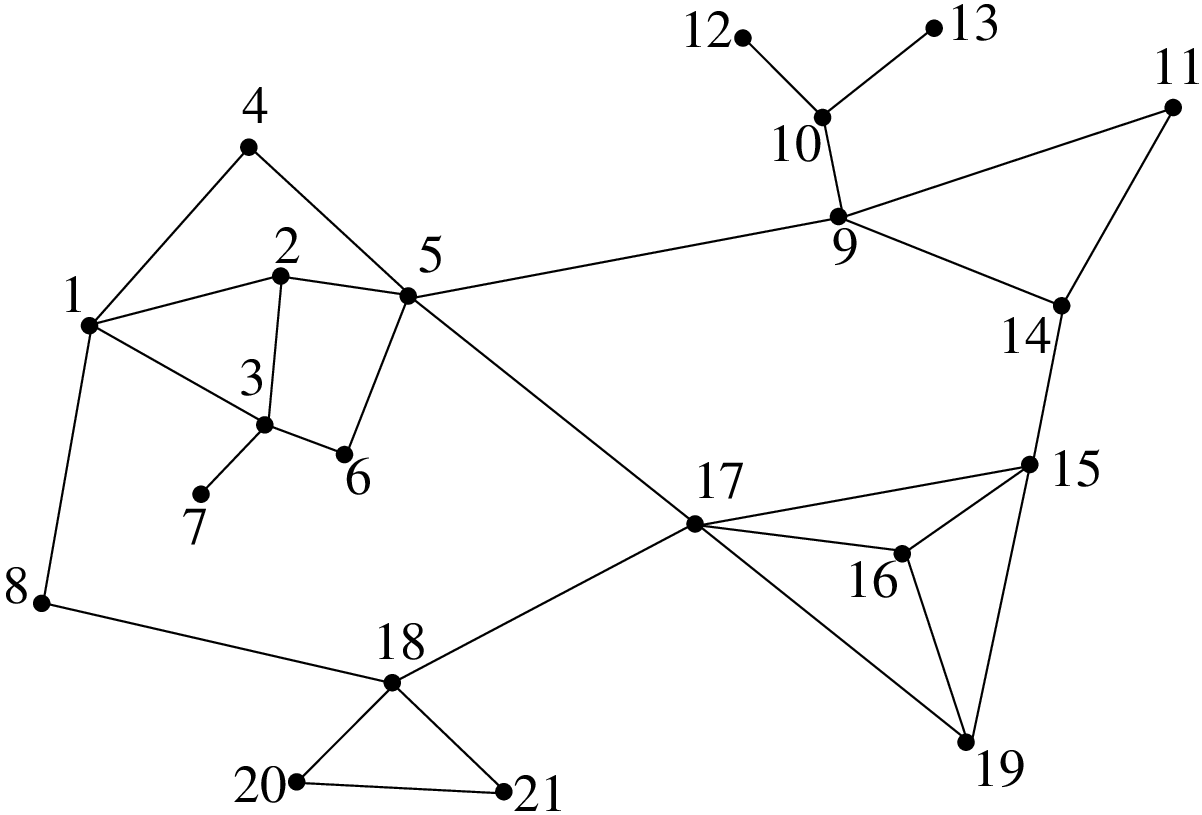}}
\caption{\label{fig:NewPic} A modular unipartite network with 21 nodes.}
\end{figure}

The interpretation of the eigenvectors of the adjacency matrix as ``community modes" is best understood as generalizing the definition of the eigenvector centrality: the eigenproblem $Au=\lambda u$ is interpreted as a self-consistent way of quantifying the centrality of the nodes on a network such that the centrality $u_i$ of node $i$ is proportional to the sum of the centralities of its neighbors, $\Sigma_{j=1}^N A_{ij} u_j$. Since centrality measures are assumed to be non-negative, only the eigenvector corresponding to the largest eigenvalue is used to define the classical centrality. On the other hand, if negative eigenvector elements are allowed, the negative signs can be transferred to the elements of $A$. We thus end up with two groups of nodes, all with positive centrality measures, but the centrality of one node is proportional to the sum of the centralities of the nodes from the same group that are connected to it minus the sum of the centralities of the nodes from the opposite group to which it is connected. This leads to meaningful bisections of the network.

\begin{figure}
\scalebox{0.5}[0.5]{\includegraphics{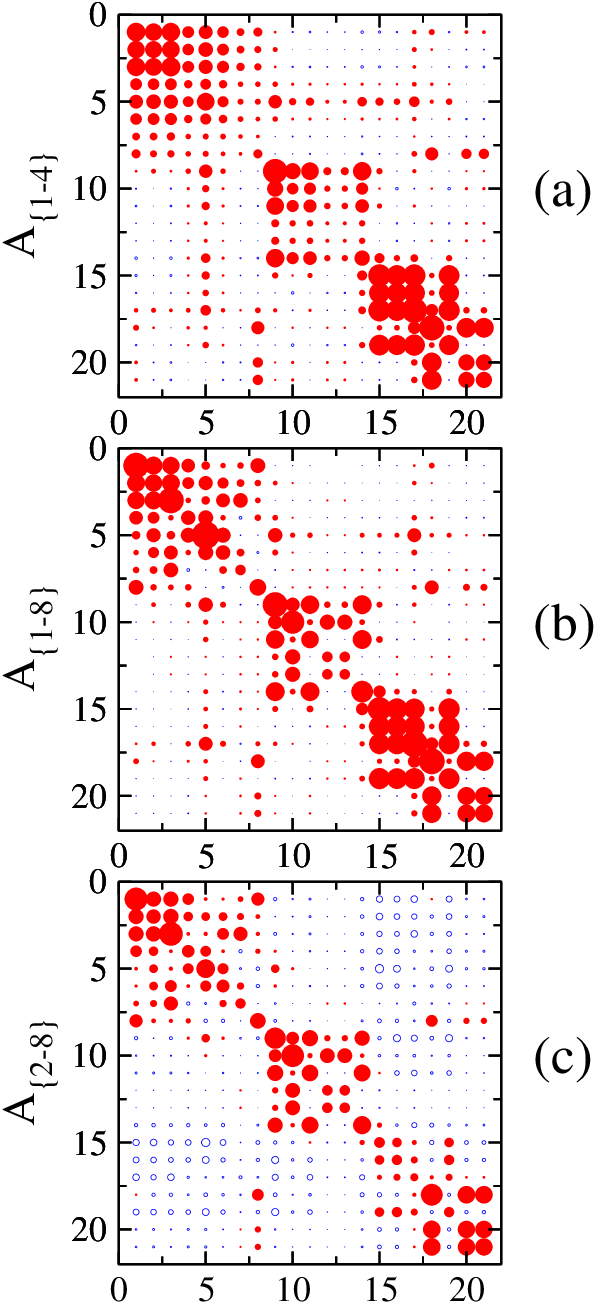}}
\caption{\label{fig:TripletNew} (Color online) Unipartite eigenvalue expansions of the adjacency matrix for the network in Fig.~\ref{fig:NewPic}. Red (solid) and blue (hollow) dots represent positive and negative matrix entries, respectively.}
\end{figure}

\subsection{Application to bipartite and directed networks}

To better understand the way the spectral split method works, let us analyze in detail what it does to a bipartite network. The eigenproblem for a bipartite adjacency matrix
\begin{equation}
A \begin{pmatrix} u \\ v \end{pmatrix} = \begin{pmatrix} 0 & B \\ B^T & 0 \end{pmatrix} \begin{pmatrix} u \\ v \end{pmatrix} = \lambda \begin{pmatrix} u \\ v \end{pmatrix}
\end{equation}
\noindent with $B$ of dimensions $m\times n$ is equivalent with
\begin{eqnarray}
(BB^T)u&=&\lambda^2 u \\
(B^TB)v&=&\lambda^2 v
\end{eqnarray}
\noindent and, if we perform a singular value decomposition
\begin{equation}
B=U\Sigma V^T ,
\end{equation}
\noindent we find
\begin{eqnarray}
BB^T&=&U\Sigma^2U^T \\
B^TB&=&V\Sigma^2V^T .
\end{eqnarray}

The eigensystem of $A$ (nullspace excluded) is thus of the form
\begin{equation}
\left\{
\pm \sigma_i,\;\frac{1}{\sqrt{2}}\begin{pmatrix} U_{:i} \\ \pm V_{:i} \end{pmatrix} \right\},\;
i=\overline{1,r}
\end{equation}
\noindent where $r \le \min(m,n)$ is the rank of $B$.

The full (non-truncated) unipartite and multipartite components of $A$ are then
\begin{eqnarray}
A_U &=& \frac{1}{2} \sum_{i=1}^r \sigma_i \begin{pmatrix} U_{:i} \\ V_{:i} \end{pmatrix}
\begin{pmatrix} U_{:i}^T & V_{:i}^T \end{pmatrix} \\
A_M &=& -\frac{1}{2} \sum_{i=1}^r \sigma_i \begin{pmatrix} U_{:i} \\ -V_{:i} \end{pmatrix}
\begin{pmatrix} U_{:i}^T & -V_{:i}^T \end{pmatrix}
\end{eqnarray}

\noindent or, in terms of $B$,
\begin{eqnarray}
A_U &=& \frac{1}{2}\begin{pmatrix} \sqrt{BB^T} & B \\ B^T & \sqrt{B^TB} \end{pmatrix} \\
A_M &=& \frac{1}{2}\begin{pmatrix} -\sqrt{BB^T} & B \\ B^T & -\sqrt{B^TB} \end{pmatrix}
\end{eqnarray}

\noindent where $\sqrt{M}$ denotes the principal, positive semi-definite root of a positive semi-definite matrix $M$.

The elements of matrices $B B^T$ and $B^T B$ count the number of ways one can travel in two steps from a node in one party to another (or the same) node in the same party. The roots of these matrices act as substitutes for the absent intraparty connections, and their low-rank approximations highlight the sets of nodes that are similarly connected in this way. Bipartite communities appear as negative entries in $A_M$.

The low-rank approximations of the unipartite component additionally highlight similar connections from either side to the other, and nodes from one party together with those from the other party through which they are connected are placed in the same community.

Note that, especially when the bipartite adjacency matrix is not written in the standard form of Eq.~(9), the best way to reveal the bipartite communities is to use

\begin{equation}
A_U-A_M=\begin{pmatrix} \sqrt{BB^T} & 0 \\ 0 & \sqrt{B^TB} \end{pmatrix} .
\end{equation}

\noindent instead of $A_M$. This prevents the off-diagonal blocks in Eq.~(19) from interfering with the bipartite community detection process and also reveals these communities through positive entries, as can be seen in Fig.~\ref{fig:net10res} (d).

In the case of directed networks, the asymmetric adjacency matrix plays the role of $B$ \cite{GuimeraModBip}. Bipartite communities are defined by similarity of only incoming or only outgoing links, whereas unipartite communities are defined based on similarity on either side and also contain the nodes to which the similar connections are made.

\subsection{A modularity-type matrix}

Discarding the first term of the unipartite component $A_U$ can be useful for revealing high-modularity unipartite community structures, which are also less likely to exhibit overlaps. This is because the matrix

\begin{equation}
A_{\{2-N\}} = A - \lambda_1 U_{:1} U_{:1}^T
\end{equation}

\noindent has similar properties with the modularity matrix defined in Eq.~(3). Since the components of $U_{:1}$ are the eigenvector centralities of the nodes, they are expected to be fairly correlated with the node degrees. Matrix $A_{\{2-N\}}$ is, in fact, a modularity-type matrix with a different null model, which uses the eigenvector centralities instead of the degrees, and $A_U-\lambda_1 U_{:1} U_{:1}^T$ is its unipartite component. The matrix depicted in Fig.~\ref{fig:net10res} (c) represents $A_{\{2-3\}}$ for the network in Fig.~\ref{fig:net10pics}.

In light of the meaning of the first term in Eq.~(5) as an outer product of the classical centrality eigenvector and best rank-1 approximation of the adjacency matrix, we see that $A_{\{1-K\}}$ provides more information about the importance of the nodes and links on the network as a whole, while $A_{\{2-K\}}$ is more focused on distinct communities, the importance of the nodes and links within them, and the possible antagonism between them. It should be noted, however, that keeping the first term does help with the detection of overlapping communities.

\subsection{Example network}

For the network in Fig.~\ref{fig:net10pics}, the truncated unipartite component of the adjacency matrix $A_{\{1-3\}}$ shown in Fig.~\ref{fig:net10res} (a) reveals three communities, comprising nodes \{1-4\}, \{5-8\} and \{7, 9, 10\}. This is consistent with the visual analysis of the network, which suggests the overlap between the latter two communities. Moreover, the importance of the ``gateway" link between nodes 3 and 5 as well as the central importance of node 7 are clearly indicated. Other smaller but significant entries indicate the stronger relationship between node 3 and nodes \{6, 8\} as well as between node 5 and nodes \{2, 4\}. Finally, the relatively close interaction between sets \{6, 8\} and \{9, 10\} is also indicated.

The modularity-type matrix $A_{\{2-3\}}$ is shown in Fig.~\ref{fig:net10res} (c). In agreement with the discussion form the previous subsection, this matrix shows non-overlapping communities \{1-4\}, \{5, 6, 8\} and \{7, 9, 10\}. These non-overlapping versions are not so well defined, presumably because of their competing tendencies to include node 7. The antagonism between sets \{3, 5\} and \{7, 9, 10\}, which tend to split the set \{5-8\} in opposite directions, is also revealed.

Figures \ref{fig:net10res} (b) and (d) reveal ``bipartite" communities \{1, 3\}, \{2, 4\}, \{5, 7\} and \{6, 8\} defined based on similarity of connection. These figures show nodes 9 and 10 each in a community by itself. This is an indication that the bipartite division of the network fails due to the link between them, with the algorithm providing an exact \emph{quadri-partite} division instead: \{1, 3, 6, 8\}, \{2, 4, 5, 7\}, \{9\}, and \{10\}, with the first two parties divided into two communities each.

For sufficiently small networks, up to about 100 nodes, the community structure can be detected by visual inspection of the truncated unipartite and multipartite components of $A$. For larger networks, two more ingredients are needed in order to have an algorithm that can automatically produce near-optimal community structures. The first is a rule for choosing the number of eigenvalues $K$. The second is an algorithm to assign the nodes to communities.

\subsection{Choosing the eigenvalue threshold}

The important structural features of a network are revealed by the most prominent positive or negative eigenvalues of its adjacency matrix and their corresponding eigenvectors. The spectra of all modular graphs examined exhibit (at least at the positive end, if no bipartite structure is discernible) a few prominent eigenvalues separated by one or more large eigengaps from the rest. This is reminiscent of properties observed in the spectrum of the unsigned Laplacian matrix \cite{SarkarSVD}. An example for a well-known network, which is discussed in detail in the Results section, is shown in Fig.~\ref{fig:EWkarate}. Numerical experiments show that the highest modularity partitions are obtained if exactly these eigenvalues are used to approximate $A_U$ or $A_M$.

However, it is important to emphasize that retaining more eigenvalues can be very useful, shedding additional light on the interactions between the nodes, despite the fact that if more eigenvalues are used to partition the network into communities the modularity will be lower. This ability to do a more in-depth analysis of the network structure is an advantage that the spectral split method offers over all community detection methods proposed thus far. Additional research, using methods similar to those described in Refs.~\cite{LanciLim, ArenasLim, FortuLim, Decelle, Nadakuditi, RadicchiPRE, RadicchiEPL}, will be required to quantify its resolution limit.

A simple rule that can be used to automatically generate high modularity community structures is to choose the threshold at the rightmost (or leftmost, in the case of the bipartite component) of the three most prominent eigengaps. More sophisticated algorithms can be devised to identify all significant eigengaps but, at least for networks of size up to $N=1000$, such algorithms seem unnecessary.

The fact that the eigenvalues separated by large eigengaps are sufficient to define the community structure is important from a computational point of view. It is known \cite{GolubBook, FortuRev} that the eigenvalues from both ends of the spectrum of a symmetric matrix and the corresponding eigenvectors can be computed by using the Lanczos algorithm \cite{Lanczos} much faster than the $\mathcal{O}(N^3)$ time required to compute the complete set of eigenvectors if these extremal eigenvalues are separated from the rest by large eigengaps.

\subsection{Assigning the nodes to communities}

The following algorithm gives good high-modularity non-overlapping partitions once a low-rank approximation of $A_U$ is computed:

\begin{enumerate} {

\item Set the negative entries of $A_{\{2-K\}}$ to zero.

\item Perform a second eigenvalue decomposition of the resulting matrix, which has only a few large, positive, eigenvalues with eigenvectors whose positive components are typically much larger than the negative ones.

\item Assume that each eigenvector corresponding to a large eigenvalue represents a community and assign each node corresponding to a positive component to that community, with a strength of the tie equal to the value of the component.

\item If non-overlapping communities are desired, assign each node to the community to which it is connected with the highest strength. For equal strengths, assign the node to the largest of the communities.

} \end{enumerate}

It is important to point out that this is just one of many algorithms that could be devised to convert the information provided by the spectral split method into community assignments. It is quite possible that other, faster and better performing, algorithms will be found.

As currently implemented by the author, with two eigenvalue decompositions and without the benefit of the Lanczos algorithm, the spectral split method can be characterized as ``intermediately fast". It is significantly faster than simulated annealing or extremal optimization, which were the two most accurate community detection methods known until now, but slower than the other, less accurate, methods mentioned in Introduction. However, the results presented in Section III show that spectral split vastly outperforms the faster methods and that it outperforms even extremal optimization in the case of large or highly modular networks. Moreover, using the Lanczos algorithm is expected to result in significant time savings, as discussed in the previous subsection.

For the purpose of comparison, the spectral split method combined with this algorithm was also applied to the classical modularity matrix $M$. Note that, in light of the discussion below Eq.~(21), it is meaningless to talk about discarding the first term in the eigenvalue expansion of $M$, and therefore $M_{\{1-K\}}$ replaces $A_{\{2-K\}}$ in this case.

Finally, a refinement that leads to small increases in modularity on some networks is to cube the eigenvalues and construct $A^3_{\{2-K\}}$ or $M^3_{\{1-K\}}$ instead of $A_{\{2-K\}}$ or $M_{\{1-K\}}$. This refinement enhances the contrast between communities defined by close eigenvalues and, even though the improvement is modest, has been used to generate the results obtained in Figs.~\ref{fig:Bres300K8}, \ref{fig:Bres300K20} and \ref{fig:Bres1000K20}.

\section{Results}

We start by presenting results for the larger modular network in Fig.~\ref{fig:NewPic}, which exhibits more features.

Figure \ref{fig:TripletNew} (a) shows the low-rank approximation $A_{\{1-4\}}$ based on the four prominent eigenvalues separated by large eigengaps from the others. In the upper-left corner there is a community consisting primarily of nodes \{1-6\}, but including nodes 7 and 8 as well. The central importance of nodes \{1-3, 5\} is clearly indicated, with node 5 highlighted as an important gateway node also connected to communities \{9-11, 14\} and \{15-17, 19\}. Node 8, though not an important member of this community, appears as a gateway node towards community \{18, 20, 21\} to which it has stronger ties. Proceeding further down along the main diagonal, we find community \{9-11, 14\} with secondary nodes 12 and 13 attached to it and then the strong communities \{15-17, 19\} and \{18, 20, 21\}. The central importance of the pairs \{14, 15\} and \{17, 18\} as gateway nodes is also highlighted by significant off-community entries.

\begin{table}
\caption{\label{tab:modularities} Comparison of the modularity values obtained for a few well-known benchmark networks.}
\begin{ruledtabular}
\begin{tabular}{ccccccc}
Network & N & $<d>$ & lev & ss(A) & ss(M) & Best \\
\hline
Karate & 34 & 4.59 & 0.3934 & 0.4174 & 0.4174 & 0.4197 \\
Dolphins & 62 & 5.13 & 0.4912 & 0.5190 & 0.5144 & 0.5285 \\
Lesmis & 77 & 6.60 & 0.5323 & 0.5526 & 0.5469 & 0.5600 \\
Football & 115 & 10.7 & 0.4926 & 0.5889 & 0.5817 & 0.6046 \\
Jazz & 198 & 27.7 & 0.3936 & 0.4328 & 0.4402 & 0.4450 \\
C. elegans & 453 & 8.97 & 0.3474 & 0.3394 & 0.3394 & 0.4520 \\
\end{tabular}
\end{ruledtabular}
\end{table}

The full-rank unipartite component $A_U=A_{\{1-8\}}$ is shown in Fig.~\ref{fig:TripletNew} (b). As expected, the additional terms included in Eq.~(7) provide more detailed information about the importance of the nodes and of the links between them. The importance of the nodes can be inferred from the diagonal elements of the matrix and the importance of the links from the off-diagonal elements. For example, within the first community, the importance of node 5 as a hub is emphasized in a way that distinguishes it from nodes \{1-3\}. Its connections with nodes 2, 4, 6, 9 and 17 are more clearly emphasized. The second community is resolved into two, \{9, 11, 14\} and \{10, 12, 13\}, with the link between 9 and 10 highlighted as an important gateway. A more detailed analysis is left to the reader, but it is clear that looking at a high-rank approximation or at the full-rank unipartite matrix provides a much richer picture of the network's structure than a simple partition into communities.

\begin{figure}
\scalebox{0.35}[0.35]{\includegraphics{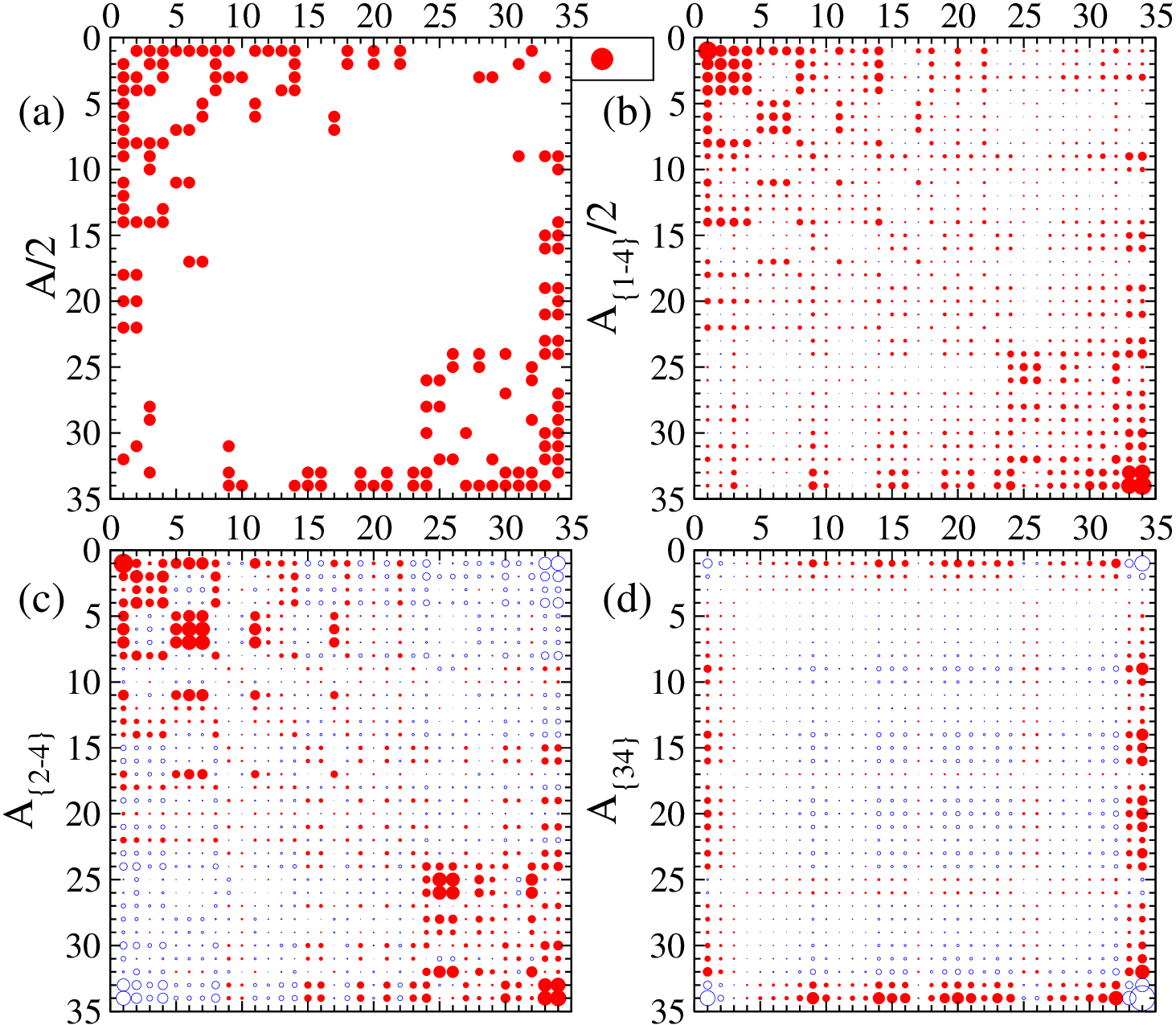}}
\caption{\label{fig:Karate-res} (Color online) The adjacency matrix and three low-rank unipartite and bipartite components for Zachary's karate network. Red (solid) and blue (hollow) dots represent positive and negative matrix entries, respectively. The dot in the legend box has unity diameter.}
\end{figure}

\begin{figure}
\scalebox{0.35}[0.35]{\includegraphics{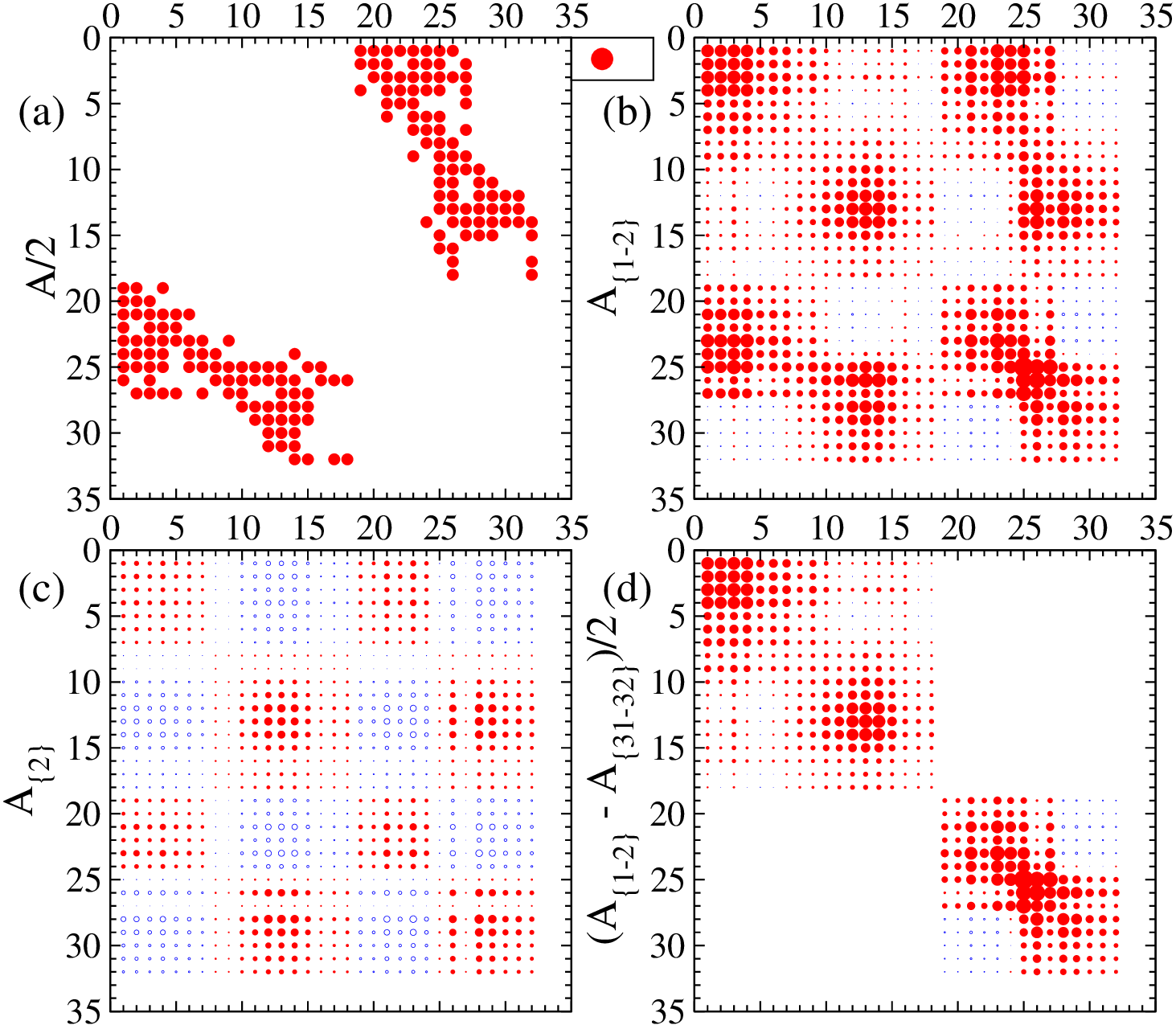}}
\caption{\label{fig:SoWo-res} (Color online) The adjacency matrix and three low-rank unipartite and bipartite components for the Southern women network. Red (solid) and blue (hollow) dots represent positive and negative matrix entries, respectively. The dot in the legend box has unity diameter.}
\end{figure}

Finally, matrix $A_{\{2-8\}}$ shown Fig.~\ref{fig:TripletNew} (c) highlights the antagonism between nodes \{1-6\} from the first community and community \{15-17, 19\}, as well as between the latter and community \{9-11, 14\}.

Detailed results for two well-known benchmark networks, the unipartite karate network of Zachary \cite{Zachary} and the bipartite Southern women network \cite{SoWo1, SoWo2} are presented next.

\subsection{Zachary's karate network}

The adjacency matrix for the karate network is shown in Fig.~\ref{fig:Karate-res} (a) and its eigenvalues in Fig.~\ref{fig:EWkarate}. The four positive eigenvalues separated by large eigengaps from the others are the ones that define a high modularity community structure. The non-overlapping partition with the maximum modularity for this network is \{1-4, 8, 12-14, 18, 20, 22\}, \{5-7, 11, 17\}, \{9, 10, 15, 16, 19, 21, 23, 27, 30, 31, 33, 34\} and \{24-26, 28, 29, 32\}, for which the Newman modularity is $Q_{max}=0.419790$. A quick inspection of Figs.~\ref{fig:Karate-res} (b) or (c) reveals a slightly different result, with an overlap between the first two communities at node 1 and an overlap between the last two communities at node 24. Both of these overlaps make sense in light of the way nodes 1 and 24 are connected. If the algorithm described in the previous section is used to generate a non-overlapping community structure, the maximum modularity partition described above is reproduced with the exception of node 24 being assigned to the third community, which results in a very slight drop in modularity to $Q=0.417406$. Note though that node 24 is connected to only two nodes in the community where it is placed by maximizing modularity and to three nodes in the community where it is placed by the spectral split algorithm.

\begin{figure}
\scalebox{0.35}[0.35]{\includegraphics{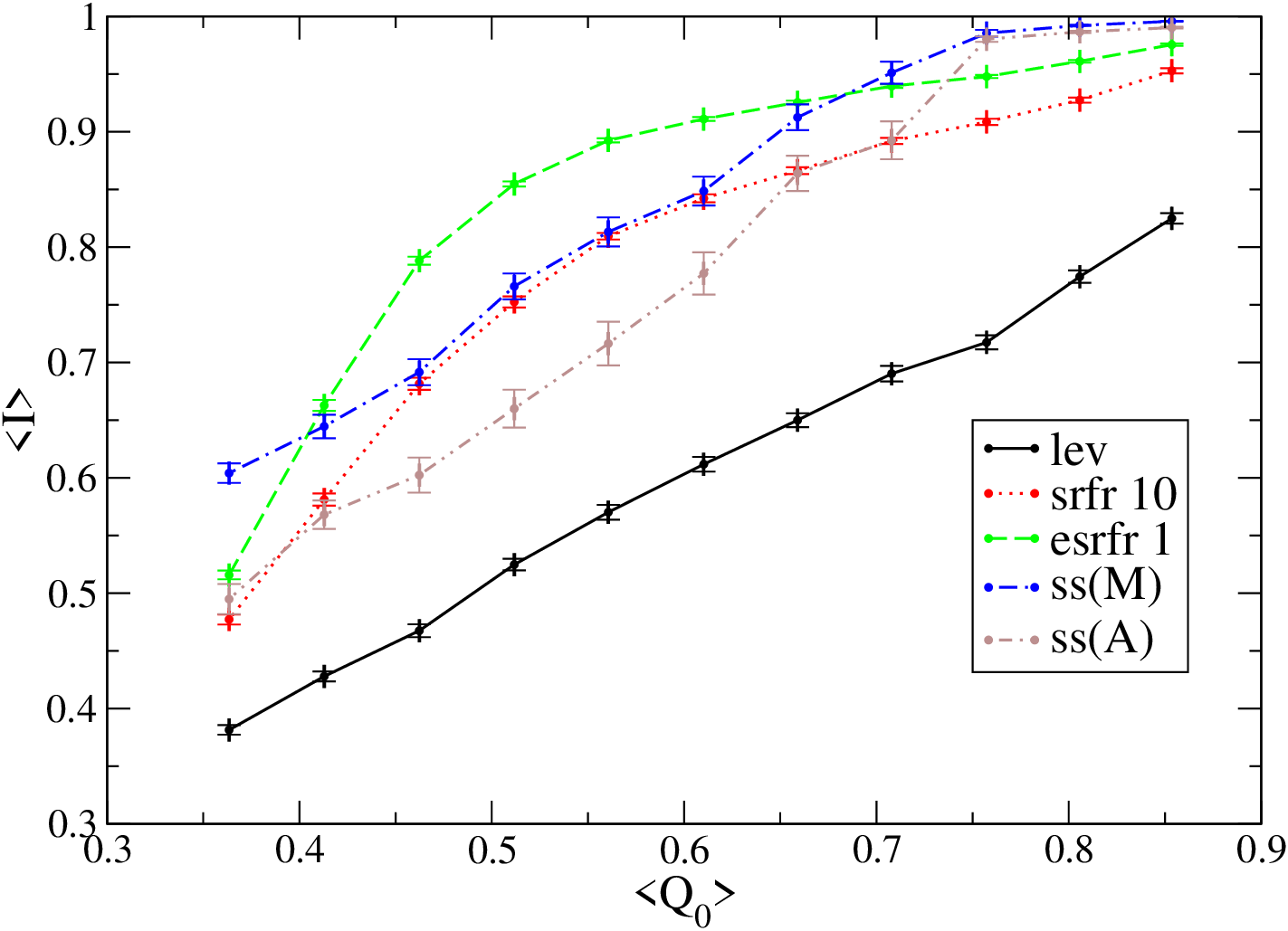}}
\caption{\label{fig:Bres300K8} (Color online) Ensemble averages of the mutual information versus the average modularity of the built-in partition for $N=300$, $<k>=8$, $k_{max}=16$. Results are presented for the leading eigenvector algorithm (unrefined: continuous black line, with refining: dotted red line), extremal optimization with refining (dashed green line), spectral split of $M$ (dash-dotted blue line), and spectral split of $A$ (dash-dot-dotted brown line).}
\end{figure}

Finally, Fig.~\ref{fig:Karate-res} (d) shows a rank-1 approximation of the bipartite component of the adjacency matrix, namely the term corresponding to the most prominent negative eigenvalue. This splits the network with nodes \{1, 2, 3, 17, 25, 26, 33, 34\} in one community and the rest of them in another, which is roughly the two opposite centers of power connected through the other nodes.

\subsection{The Southern women network}

This network is the most frequently used benchmark for bipartite community detection algorithms \cite{GuimeraModBip, BarberModBip, SarkarSVD}. Nodes 1 through 18 represent women, while nodes 19 through 32 represent events in which they participated. The original partition into communities, given by the authors of Ref.~\cite{SoWo1}, pertains only to women and is an overlapping one: \{1-9\} and \{9-18\}. The adjacency matrix for this network is shown in Fig.~\ref{fig:SoWo-res} (a) while unipartite and bipartite components for $K=2$ are shown in Figs.~\ref{fig:SoWo-res} (b-d).

By inspection of Fig.~\ref{fig:SoWo-res} (b) we find overlapping unipartite communities \{1-10, 19-27\} and \{3, 7-18, 25-32\} while Fig.~\ref{fig:SoWo-res} (d) reveals overlapping bipartite communities \{1-10\}, \{3,7-18\}, \{19-27\} and \{25-32\}. A more careful consideration of the link weights shows that the only significant overlaps between the women communities occur at nodes 8 and 9, which is in good agreement with the original partition. Note that in this simple case, where the network is rigorously bipartite and divided using very low-rank approximations of the adjacency matrix, the bipartite communities can be expressed as intersections between the unipartite communities and either party. This is not necessarily the case, however, if higher-rank approximations of the adjacency matrix are used or if the network is only approximately bipartite.

\begin{figure}
\scalebox{0.35}[0.35]{\includegraphics{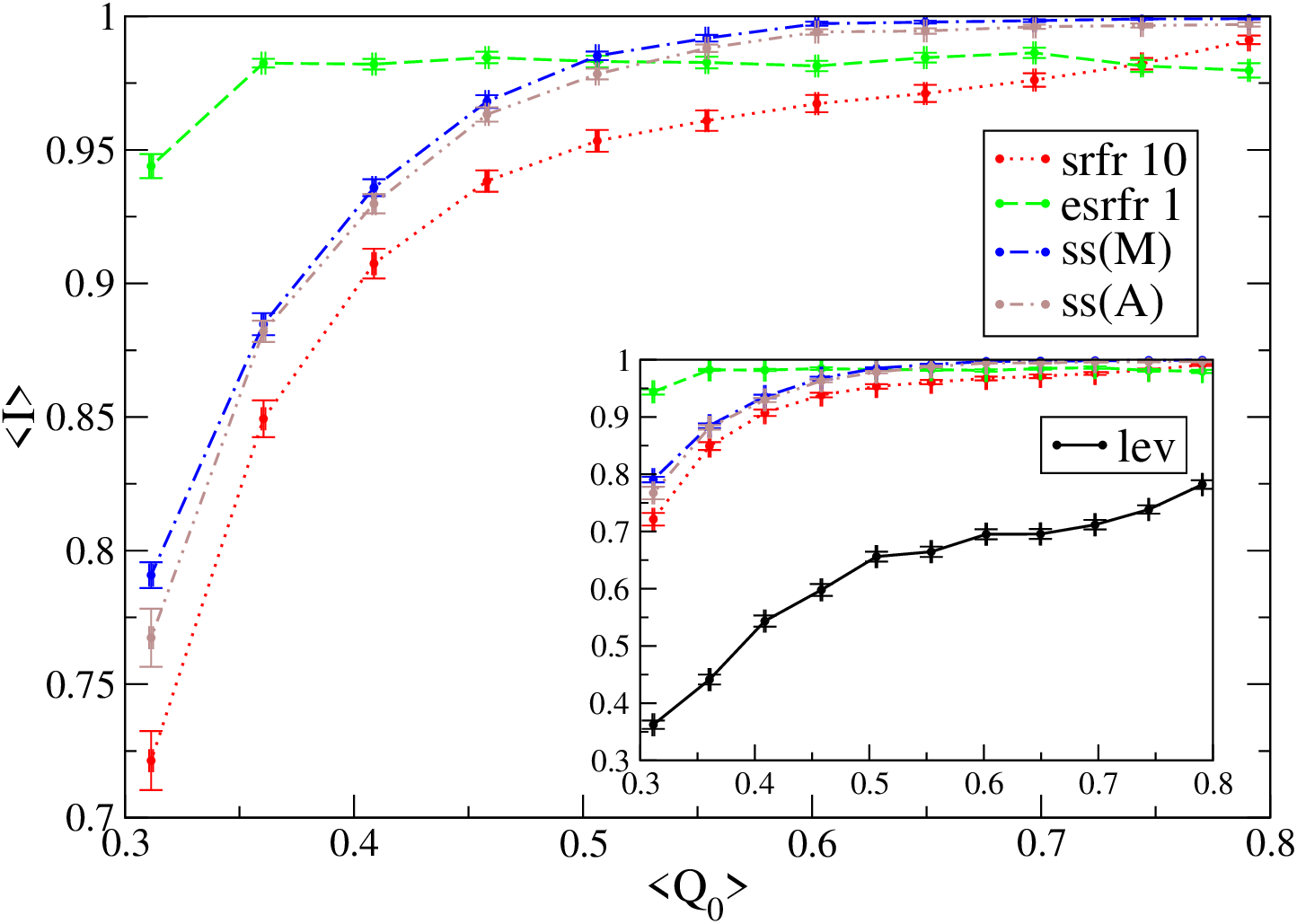}}
\caption{\label{fig:Bres300K20} (Color online) Ensemble averages of the mutual information versus the average modularity of the built-in partition for $N=300$, $<k>=20$, $k_{max}=40$. Results are presented for the leading eigenvector algorithm (unrefined: continuous black line, with refining: dotted red line), extremal optimization with refining (dashed green line), spectral split of $M$ (dash-dotted blue line), and spectral split of $A$ (dash-dot-dotted brown line).}
\end{figure}

Matrix $A_{\{2\}}$, which is depicted in Fig.~\ref{fig:SoWo-res} (c), reveals two unipartite non-overlapping communities: \{1-7, 19-24\} and \{8-18, 25-32\}. This result is very close to the partition obtained in Refs.~\cite{GuimeraModBip, BarberModBip} for the case of division into two communities, namely \{1-7, 9, 19-26\} and \{8, 10-18, 27-32\}.

With regard to the bipartite communities, the highest modularity division reported in Ref.~\cite{BarberModBip} is \{1-6\}, \{7,9,10\}, \{8,16-18\}, \{11-15\}, \{19-24\}, \{25,26\}, \{27,29\} and \{28,30-32\}. Similar partitions can be obtained with the spectral split algorithm if more eigenvalues are included. For example, using $A_{\{1-3\}}-A_{\{30-32\}}$ we find partitions \{1-7, 9, 10\}, \{8, 16-18\}, \{11-15\}, \{19-25, 27\}, \{26\} and \{28-32\}.

Finally, Figs.~\ref{fig:SoWo-res} (b) and (d) also show the higher importance of nodes \{25-27\}, which represent events \{7-9\} and were attended by many women from both groups \cite{SoWo1, SarkarSVD}. The event communities are actually shown to be overlapped at these nodes.

\subsection{Other benchmark networks}

Table~\ref{tab:modularities} shows a comparison of the modularities obtained using the spectral split method applied to the adjacency matrix and to the modularity matrix, denoted by $ss(A)$ and $ss(M)$, respectively, with those obtained using the unrefined leading eigenvector method \cite{NewmanPNAS}, denoted by $lev$, and with the highest modularity results found in literature \cite{HoustonGang}. Included are some of the best-known networks, namely Zachary's karate network \cite{Zachary}, the dolphins network of Lusseau et al., the network of interactions between the characters in Victor Hugo's ``Les Miserables" \cite{LesMis}, the American college football network first studied by Girvan and Newman \cite{Football}, the network of jazz musicians \cite{Jazz}, and the metabolic network of the worm C. elegans \cite{CelegMeta}.

With the exception of the C. elegans metabolic network, both applications of the spectral split algorithm compare very well with the other methods, and $ss(A)$ seems generally better than $ss(M)$. Note that the highest modularity results are typically obtained by simulated annealing or extremal optimization, which are much slower methods. The results in Table~\ref{tab:modularities} suggest that the spectral split method works better for networks with higher average degree or higher modularity. They also seem to hint that the algorithm might not work well for larger networks.

\subsection{Statistical ensemble results}

To check the validity of these statements and to quantify the performance of the algorithm, tests were performed on ensembles of random benchmark networks generated using the algorithm from Ref.~\cite{Benchmarks}. These are scale-free networks with a built-in community structure. They have a number of tunable parameters, which include the average degree, the maximum degree, and the mixing parameter $\mu$, which represents the average fraction of links running between different modules and controls the average modularity of the statistical ensemble of networks. The parameters not discussed here were kept at their default values.

\begin{figure}
\scalebox{0.35}[0.35]{\includegraphics{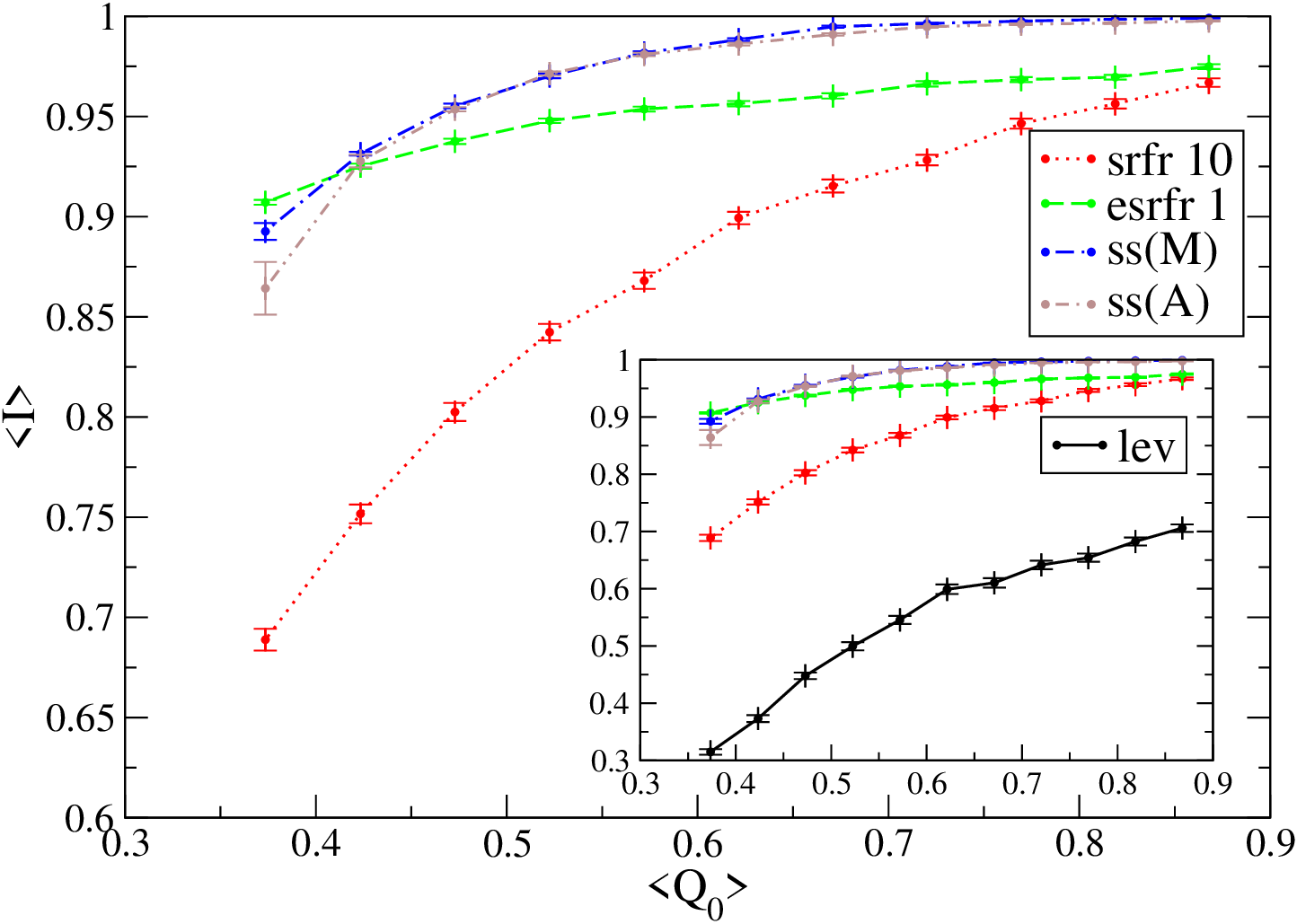}}
\caption{\label{fig:Bres1000K20} (Color online) Ensemble averages of the mutual information versus the average modularity of the built-in partition for $N=1000$, $<k>=20$, $k_{max}=40$. Results are presented for the leading eigenvector algorithm (unrefined: continuous black line, with refining: dotted red line), extremal optimization with refining (dashed green line), spectral split of $M$ (dash-dotted blue line), and spectral split of $A$ (dash-dot-dotted brown line).}
\end{figure}

Tests were performed on networks of size $N$ between 100 and 1000, average degree $\left<d\right>$ between 6 and 30 and maximum degree up to 100. Some of the results are presented in Figs.~\ref{fig:Bres300K8}, \ref{fig:Bres300K20}, and \ref{fig:Bres1000K20}. The data points in these figures represent averages computed over ensembles of 100 networks with fixed values of the mixing parameter $\mu$. The average mutual information between the computed and the built-in partitions is plotted versus the average modularity of the built-in partition. The error bars represent the standard error of the mean. To obtain the different points, $\mu$ was varied between 0.1 and 0.6 in steps of 0.05.

The spectral split method [both $ss(A)$ and $ss(M)$] is compared with three other methods implemented using the \texttt{Radatools} software package \cite{Radatools}. These are the leading eigenvector method \cite{NewmanPNAS} without refining ($lev$), the same method with multiple Kernighan-Lin-like and greedy optimization refining \cite{NewmanPNAS, NewmanGreedy} repeated 10 times (heuristics string \texttt{srfr 10}), and the extremal optimization method of Ref.~\cite{ExtrOpt} followed by spectral optimization and refining (heuristics string \texttt{esrfr 1}).

It is clear that increasing network size does not reduce the ability of the spectral split method to detect the correct community structure. Quite to the contrary, it is in the case of large networks that it compares most favorably with its peers. Note that the $N=300$ and $N=1000$ networks from the high-modularity ensembles routinely exhibit 10 to 20 communities. Spectral split is vastly superior to the unrefined leading eigenvector method, and it overtakes all the other methods, including extremal optimization, in the case of networks with significant modularity.

On the other hand it is true that, without refinement, the spectral split algorithm falls behind extremal optimization in the case of low-modularity or very sparse networks. For networks that are not very sparse, the low values of modularity at which this happens are comparable to those of similar random networks, and therefore it is questionable whether such community structure is truly meaningful \cite{FortuRev}.

In regards to speed we note that, although slower than less accurate methods, spectral split is faster than extremal optimization or simulated annealing while offering comparable accuracy. For example, in the case of networks of size $N=1000$ it is an order of magnitude faster than extremal optimization even without using the Lanczos algorithm to compute the eigenpairs.

Finally, $ss(M)$ appears superior to $ss(A)$ on very sparse networks, but the difference in performance between the two variants is negligible in all other cases and decreases with increasing network size. If we also consider the results obtained in the previous subsection, which show $ss(A)$ outperforming $ss(M)$ on real-world networks, we conclude that the comparison between them is probably a complex issue that depends on many aspects of network topology.

\section{Conclusions}

A new method for analyzing the structure of complex networks was introduced. This method does more than simply partition the network into communities, providing information, at different levels of detail, about the strengths of the interactions between the nodes. In this regard, it is useful even without an actual grouping of the nodes into communities. The spectral split method introduced in this paper can be applied to the adjacency matrix, in which case it can reveal both unipartite and bipartite community structures, but for unipartite networks it can also be applied to the modularity matrix. An algorithm is also introduced for the purpose of constructing the communities. Tests on statistical ensembles of benchmark networks show that the spectral split method combined with this algorithm produces excellent results, especially in the case of large networks or networks with significant modularity. It is possible that further research will produce faster and better-performing community assignment algorithms which will make the spectral split method even more competitive.

\bibliography{Paper4}

\end{document}